\documentclass[preprint2]{aastex}
\usepackage{subfigure}
\usepackage{enumerate}
\usepackage{courier}
\usepackage{epsfig}
\usepackage{lscape}
\usepackage{color,soul}
\usepackage[normalem]{ulem}
\sethlcolor{yellow}
\usepackage{amsmath}

\shorttitle{Elemental Abundance Differences in WASP-94AB}
\shortauthors{Teske, Khanal, \& Ram\'irez}

\begin{document}
\sethlcolor{yellow}

\newcommand{\txw}{\textwidth}

\title{The Curious Case of Elemental Abundance Differences in the Dual Hot Jupiter Hosts WASP-94AB$^{*}$}

\altaffiltext{*}{This paper includes data gathered with the 6.5 meter Magellan Telescopes located at Las Campanas Observatory, Chile.}

\author{Johanna K. Teske\altaffilmark{1,+}, Sandhya
  Khanal\altaffilmark{2}, Ivan Ram\'irez\altaffilmark{2}}

\altaffiltext{1}{Carnegie Department of Terrestrial Magnetism; 5241 Broad Branch Road, NW, Washington, DC 20015, email: jteske@carnegiescience.edu}
\altaffiltext{2}{McDonald Observatory and Department of Astronomy, University of Texas at Austin; 2515 Speedway, Stop C1402, Austin, TX 78712-1205, USA}
\altaffiltext{+}{Carnegie Origins Fellow, jointly appointed by Carnegie DTM \& Carnegie Observatories}

\begin{abstract}

Binary stars provide an ideal laboratory for investigating the
potential effects of planet formation on stellar composition. Assuming the stars formed in the
same environment/from the same material, any compositional anomalies
between binary components might indicate differences in how
material was sequestered in planets, or accreted by the star in the
process of planet formation. We present here a study of the elemental abundance
differences between WASP-94AB, a pair of stars that each host a hot Jupiter exoplanet. The two stars are very similar in
spectral type (F8 and F9), and their $\sim$2700 AU separation suggests their
protoplanetary disks were likely not influenced by stellar
interactions, but WASP-94Ab's orbit -- misaligned with the host star
spin axis and likely retrograde -- points towards a
dynamically active formation mechanism, perhaps different than that of
WASP-94Bb, which is not misaligned and has nearly circular orbit. Based on our
high-quality spectra and strictly relative abundance analysis, we
detect a depletion of volatiles ($\sim$-0.02 dex, on average) and enhancement of
refractories ($\sim$0.01 dex) in
WASP-94A relative to B (standard errors are $\sim$0.005 dex). This is different than every other published
case of binary host star abundances, in which either no significant
abundance differences are reported, or there is some degree of enhancement in
all elements, including volatiles. Several scenarios that may explain
the abundance trend are discussed, but none can be definitively accepted or
rejected. Additional high-contrast imaging observations to search for
companions that may be dynamically affecting the system, as well as a
larger sample of binary host star studies, are needed to
better understand the curious abundance trends we observe in WASP-94AB.

\end{abstract}

\keywords{planets and satellites: formation --- planets and satellites: individual (WASP-94) --- stars: abundances --- stars: atmospheres}

\section{Introduction}
In the study of exoplanets, host star composition is of interest
because it may serve as a proxy for planetary
composition. Early on, this connection was viewed from a
``star-centric'' perspective -- how does the
formation and evolution of planets change the original, pre-planet
composition of stars? Subsequently, once the correlation between the presence of
giant planets and host star metallicity was established as primordial
(e.g., Santos et
al.\,2004; Fischer \& Valenti 2005), a more ``planet-centric'' perspective was adopted -- how does host star
composition influence the type of planets that form? While many
investigations have focused on the latter perspective, seeking to
expand upon the observed giant planet-metallicity correlation (e.g., Sousa et al. 2008; 
Ghezzi et al. 2010; Adibekyan et al. 2012; Everett et al. 2013; da
Silva et al. 2015; Buchhave \& Latham 2015), a new growing body of work aims
to revist the ``star-centric'' perspective by examining differences
between very similar stars that do/do not host known planets (e.g.,
Mel\'endez et al. 2009; Schuler et al. 2011b; Ram\'irez et al.\,2011; Ram\'irez et al. 2014; Tucci Maia
et al. 2014; Liu et al. 2014; Nissen 2015; Saffe et al. 2015), or that
are known to host different types of planets (Mack et al. 2014; Teske
et al. 2013, 2015; Ram\'irez et al. 2015). 

Measuring stellar abundances with enough precision to detect
differences that may be due to planets formation is extremely
challenging. Mel\'endez et al. (2009) suggested that their measured
$\sim$20\% deficit of refractory elements (condensation temperatures $T_c\gtrsim$ 1000
K) in the Sun versus other ``solar twins'' of similar T$_{eff}$, log $g$,
and [Fe/H]\footnote{[X/H]=log(N$_{\rm{X}}$/N$_{\rm{H}}$) - log
  (N$_{\rm{X}}$/N$_{\rm{H}}$)$_{\rm{solar}}$} was due to the formation
of terrestrial planets in our Solar System. This result hinged on
abundance errors of 0.01 dex or less, a level of precision made
possible only by studying ``twin'' stars in a strictly differential
analysis, using high resolution ($R \gtrsim$ 50,000), high
signal-to-noise ($\gtrsim$ 400) spectra. The results of Mel\'endez et al. (2009) have been both
questioned (e.g., Schuler et al. 2011a; Gonz\'alez Hern\'andez et al.\,2010, 2013; Adibekyan et
al. 2014; Gaidos 2015) and replicated (e.g., Ram\'irez et
al.\, 2009, 2010; Nissen 2015); the interpretation will likely remain
suspect until a larger sample of planet host stars is
examined in similar detail. However, the potential knowledge gained from
measuring ``missing'' or ``added'' material to host stars is great -- for most exoplanets, now and in the near
future, we will only know their orbital period and their mass or
radius, meager information from which to deduce a
composition. The power of studying host star compositions to learn
about their orbiting planets is already
demonstrated by higher stellar metallicities being indicative of
higher probabilities of giant planet detections. 

Binary stars are particularly useful for studying differences in
stellar composition that may be related to the formation of planets. Like the classic experimental
setup, if one star in a binary system is known to host a planet and the other is not,
the star without a detected planet serves as the ``control'', a relic of the
primordial composition of the system (assuming the stars formed in the
same environment/from the same material) and is unlikely to be influenced by different
Galactic chemical evolution (Adibekyan et al. 2014) or by peculiar environment
effects like supernovae pollution or dust depletion by nearby hot
stars (\"{O}nehag et al. 2011). Studying
``twin'' stars ($\Delta T_{eff} \lesssim 100, \Delta$log $g \lesssim 0.1$) allows very
precise relative abundance measurements, as systematic uncertainties that
usually dominate abundance analyses are so similar that they
essentially cancel out. The remaining observational noise can then be pushed down
with very high signal-to-noise, high-resolution spectra. 
Any observed differences in
composition between the binary components then \textit{could} be related to
planet formation. Work by Desidera et al. (2004; 2006) on a large
survey of wide binaries suggested that [Fe/H] differences between
binary stars $\geq$0.03 are
atypical. Gratton et al. (2001) found that four out of six wide
binaries had indistinguishable abundances (at the $\leq$0.012 dex level) in elements ranging from
$T_c \sim 100$ to 1700 K. However, in these
previous works the binaries were not always composed of ``twins'', and the
data were not always high signal-to-noise ($\gtrsim 200$). More recent work on the true
``twin'' system XO-2 suggests that differences in stellar abundances
as small as 0.015 dex could be due to a
different type of planet formation around one star versus the other
(Ram\'irez et al. 2015; see also Teske et
al. 2015, Biazzo et al. 2015). Even this small difference amounts to
$\sim$0.57M$_J$ of volatile-rich material potentially ``missing''
from XO-2S, instead locked up in its two gas giant planets. Mel\'endez et al. (2009)'s
$\sim$0.08 dex difference in refractory element abundance between the
Sun and 11 solar twins corresponds to $\sim$4M$_{\odot}$ of rocky material, which the authors
equate to the material contained within the Solar System terrestrial planets (Chambers 2010).

In an effort to further understand how planets affect/are affected by
their host star composition, here we expand the sample of well-studied
``twin'' binary host systems, performing a high-precision compositional analysis of WASP-94AB, in which both stars are known to host a single close-in giant planet.

\section{The WASP-94 System}

The planets in the WASP-94AB system were first reported by
Neveu-VanMalle et al. (2014; NV14), identified through the WASP-South
transiting planet detection program (Hellier et al. 2011). The binary
consists of a $V$=10.1, F8V primary, separated by 15$\arcsec$ from the
$V$=10.5, F9V secondary. The consistent proper motions, radial
velocities of the stars, and little
change over time of the position angle or separation (as described in
NV14) indicate that the components are likely
bound. From the measured absolute magnitudes, NV14 estimate the
projected separation between WASP-94A and B to be $\geq$2700 AU. The
primary star hosts a transiting 0.45 M$_J$, 1.72 R$_J$ planet at 0.055 AU, while the secondary hosts a
non-transiting planet at 0.034 AU with a minimum mass $M_P~sini$ of
0.62 M$_J$. A combined analysis of WASP-94Ab transit and radial
velocity data limits an eccentricity $e < 0.13$ at the 3$\sigma$
level, but the orbit of the planet is obviously misaligned with the
stellar rotation axis, and is probably retrograde ($\lambda=~151^{\circ}
\pm 20^{\circ}$). NV14 assumes a circular orbit for WASP-94Bb, as the
error on the fitted $e$ is larger than the value itself, and their
unsuccessful search for a transit of the planet provides an estimate
of the inclination of WASP-94Bb's orbital plane of $i
\lesssim$79$^{\circ}$. 

WASP-94AB is one of only three known resolvable\footnote{Kepler-132 is
  a stellar binary that hosts three small planets, but due to the
  small angular separation it is unknown
  around which star the planets orbit (Lissauer et al. 2014). See also
Deacon et al. (2015).} stellar
``twin'' systems in which \textit{both}
stars host planets (in S type orbits, i.e., not orbiting both
stars). The HD20782/20781 binary, a G1.5V/G9.5V
pair separated by $\sim$9000 AU, is actually on the border of the colloquial ``twin'' regime, with a
$\Delta$T$_{eff} \sim 500$ K and $\Delta$log $g \sim$0.10 dex. The primary hosts a 1.9 M$_J$, 1.4 AU, $e
\sim$0.97 planet and the secondary hosts a 0.04 M$_J$, 0.17 AU, $e
\sim 0.11$ planet and a 0.05 M$_J$, 0.35 AU, $e \sim 0.28$
planet; all three of the planets were detected via radial
velocity (Jones et al. 2006; Mayor et al. 2011).
Mack et al. (2014) found no significant abundance differences
between the HD20782/20781 stars, but reported errors between 0.02-0.07
dex. The other known dual-hosting binary is
XO-2NS, a $\sim$4600 AU-separated pair of $\sim$0.97 M$_{\odot}$ stars
that are more ``twin''-like, with $\Delta$T$_{eff} \sim$60 K,
$\Delta$log $g \sim$0.02 dex, and $\Delta$[Fe/H]$\sim$0.06 dex
(Biazzo et al. 2015; Ram\'irez et al. 2015; Teske et al. 2015; Damasso
et al. 2015; Desidera et al. 2014). XO-2N hosts a transiting planet
with 0.62 M$_J$ at 0.04 AU (assumed $e=0$; Burke et al. 2007), and
XO-2S hosts two radial-velocity-detected planets, with masses 0.26
M$_J$ and 1.37 M$_J$, orbital separations 0.13 and 0.48 AU, and $e
\sim$0.18 and 0.15, respectively (Desidera et al. 2014; Damasso et
al. 2015). 

Thus, WASP-94 differs from HD20782/20781 and XO-2 in a few important
ways: 

\begin{enumerate}
\item The binary separation is smaller, at 2700 AU versus 4600 AU (XO-2)
  and 9000 AU (HD20782/20781).

\item The stars are hotter and more massive, meaning their convective
  envelopes are smaller, and changes due to planet formation may be
  easier to detect. 

\item Both stars are known to host a single low-eccentricity, close-in
  ($<$0.06 AU) gas giant planets, versus higher-eccentricity/longer-orbit/multiple
  planets.

\end{enumerate}

\noindent WASP-94 is of particular interest because hot Jupiter planets (roughly
defined as $M_P~sini >0.1 M_J$ and P$<$10 days) are rare,
found around $\lesssim$1\% of FGK dwarfs (Wright et al. 2012,
Table 2). The planets in this system must have migrated inward, but
what triggered their (and other hot Jupiter) period shrinkage is an open question, as 
there are several proposed mechanism(s) for giant planet migration in a
protoplanetary disk (e.g., Goldreich \& Tremaine 1980; Wu \& Murray 2003;
Naoz et al. 2011; Rasio \& Ford 1996; Wu \& Lithwick 2011; Guillochon
et al. 2011). Furthermore, while both planets likely
have near zero eccentricities, WASP-94Ab has a misaligned and probably
retrograde orbit, WASP-94Bb's orbit is inclined relative to Ab, and
(comparing age-based periods with periods computed via $v~sini$)
WASP-94B (the star) may be inclined by $>$60$^{\circ}$ with respect to WASP-94A (NV14). This system is an
exceptional laboratory for studying the formation and dynamical evolution of hot Jupiter exoplanets via
changes they may have induced in their host stars. 

\section{Spectroscopic Observations and Analysis}

Observations of WASP-94AB were acquired on 1 May 2015 (UT) with
the MIKE high resolution spectrograph (Bernstein et al. 2003) on the
6.5m Clay Magellan Telescope at Las Campanas Observatory. The standard
MIKE setup was used with a 0.5$\arcsec$ slit, providing
$\sim$320-1000nm wavelength coverage and $R \sim$45,000. Multiple exposures were taken of each star and added
together during the reduction process to achieve a peak $S/N$ at
6000 {\AA} of $\sim$500 for WASP-94A and $\sim$400 for WASP-94B. The
spectra were reduced with the CarnegiePython MIKE pipeline
\footnote{http://code.obs.carnegiescience.edu/mike}, then
velocity-corrected and combined with IRAF\footnote{IRAF is distributed
  by the National Optical Astronomy Observatory, which is operated by
  the Association of Universities for Research in Astronomy (AURA)
  under cooperative agreement with the National Science Foundation.} following the
procedure described in Ram\'irez et al. (2014). 

The stellar parameters and abundances detailed below were derived from
equivalent width (EW) measurements -- fitting Gaussian functions to
observed line profiles with IRAF's \texttt{splot} task -- of spectral lines, unless otherwise
noted. The line list was taken from Ram\'irez et al. (2014), which
includes Fe I lines covering a wide range of excitation potentials and
$\sim$20 Fe II lines. This line list was selected to include only
lines with strengths low enough to be on the linear part of the curve
of growth, and to exclude blended lines or those in low S/N parts of
the spectrum. We performed independent measurements of
each line (Table \ref{lines}) using \texttt{splot} in IRAF, and the resulting abundances were
averaged together for the final reported values. Independent
measurements of the stars were performed after initial measurements
suggested the $\Delta$[X/H] differences were very small and barely
detectable with only one set of EW measurements. Since the stars are ``twins'', the errors in $\Delta$[X/H]
can be reduced by minimizing the observational noise, some of which
comes from the precise way in which each person decides to measure
EWs (e.g., the continuum placement). Combining the EWs instead of the
derived parameters would have led to larger errors due to the
systematics between EW sets. The solar reference EW
measurements are also from Ram\'irez et al. (2014). 

The EW measurements were translated into elemental abundances using
the curve-of-growth method via the 2014 version of the spectral analysis code MOOG (Sneden
1973), specifically the \texttt{abfind} driver, and ``standard composition'' MARCS 1D-LTE stellar atmosphere
models (Gustafsson et al. 2008), linearly interpolated to the
appropriate stellar parameters. This is the third paper to utilize the
publicly-available \texttt{Qoyllur-quipu}
($q^2$) Python package\footnote{https://github.com/astroChasqui/q2},
a MOOG wrapper that uses EWs as input to derive stellar parameters and
abundances. Given the measured EWs in Table \ref{lines}, the results
presented here can be fully reproduced with $q^2$.

\subsection{Stellar Parameters}

Fundamental stellar parameters $T_{eff}$ (effective temperature), log
$g$ (surface gravity), [Fe/H] (iron abundance), and $\xi$
(microturbulent velocity) were determined from a traditional
equilibrium/ionization balance. In this method, the differences in Fe
I and Fe II abundances are compared to the excitation potential and reduced equivalent width ($REW$ =
log $EW/\lambda$) in a line-by-line manner to minimize any
correlations. The abundances of Fe I and Fe II depend on the input
stellar model, which is iteratively modified (in $T_{eff}$, log $g$,
[Fe/H], and $\xi$) until the
excitation/ionization balance conditions are reached. The details of
this procedure are described in Ram\'irez et al. (2014) and Teske et
al. (2014). Here, we employ the formal error determination of
Ram\'irez et al. (2014), which propagates both the measurement errors
(line-by-line variation in abundances) and uncertainties in the
derived stellar parameters (Epstein et al. 2010; Bensby et
al. 2014). In each case described below, we derived parameters from
each independent set of EW measurements, and the reported values are
the mean $\pm$ the error on the mean\footnote{Defined as
  $\sqrt{\frac{\sum_{n=1}^{n=3} x_{err}^2}{n}}$, where $x_{err}$= the
  error on each independently-derived parameter}. 


We first determined WASP-94A and -94B stellar parameters differentially with
respect to the Sun, assuming a solar $T_{eff}=$5777 K, log
$g=$4.44 dex, [Fe/H]=0, and $\xi=$1.00 km s$^{-1}$. The resulting
parameters and errors for each star are given in Table \ref{params}, as are the
originally published parameters and errors from NV14. We find slightly
higher values for $T_{eff}$, log $g$, and [Fe/H] than NV14 (see
second section Table \ref{params}), but our results
overlap within errors in every case except log $g$ for WASP-94B, where
the difference including errors is 0.04 dex. 

Unlike the solar twins in Mel\'endez et al. (2009), it is clear that WASP-94AB
are significantly different than Sun in $T_{eff}$, log $g$, and
[Fe/H]. Our solar spectra were obtained with the same instrument and
setup, but not on the same night, which is not ideal. Furthermore,
there may be small differences in the Galactic chemical evolution of the Sun versus WASP-94AB (which has space
motions consistent with young disk stars, and is 180$\pm$20 pc
away). Thus, we can obtain more precise parameters by comparing the
two stars differentially to each other; the differences between the
two stars are the desired quantities, not how they compare to the
Sun. We assume the parameters
derived above for WASP-94B as fixed, since it is closer in its parameters to
solar and thus the Sun-relative analysis is more reliable than for
-94A, and proceed to measure only the relative parameters for WASP-94A. These
results are listed in the third section of Table \ref{params}; the
WASP-94A parameters change little from our Solar Reference case, but
the errors are $\sim$halved. These strictly (A-B) differential values result in $\Delta T_{eff}$,
$\Delta$log $g$, and $\Delta$[Fe/H] values that overlap with the 
$\Delta$(A-B) parameter differences from NV14 (see third lines of each section
of Table \ref{params}). 

\begin{figure}[ht!]
\includegraphics[width=0.49\textwidth]{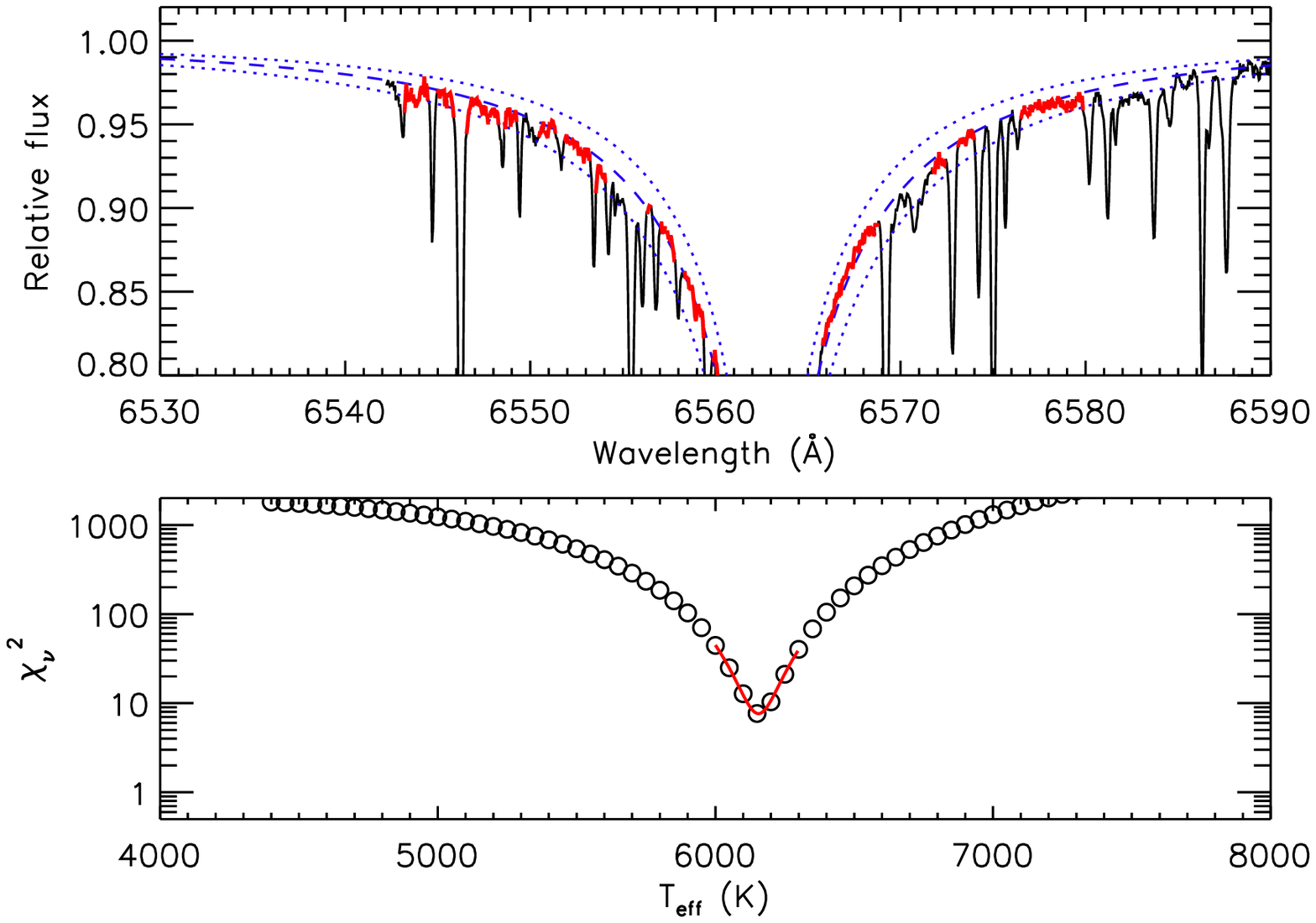}
\includegraphics[width=0.49\textwidth]{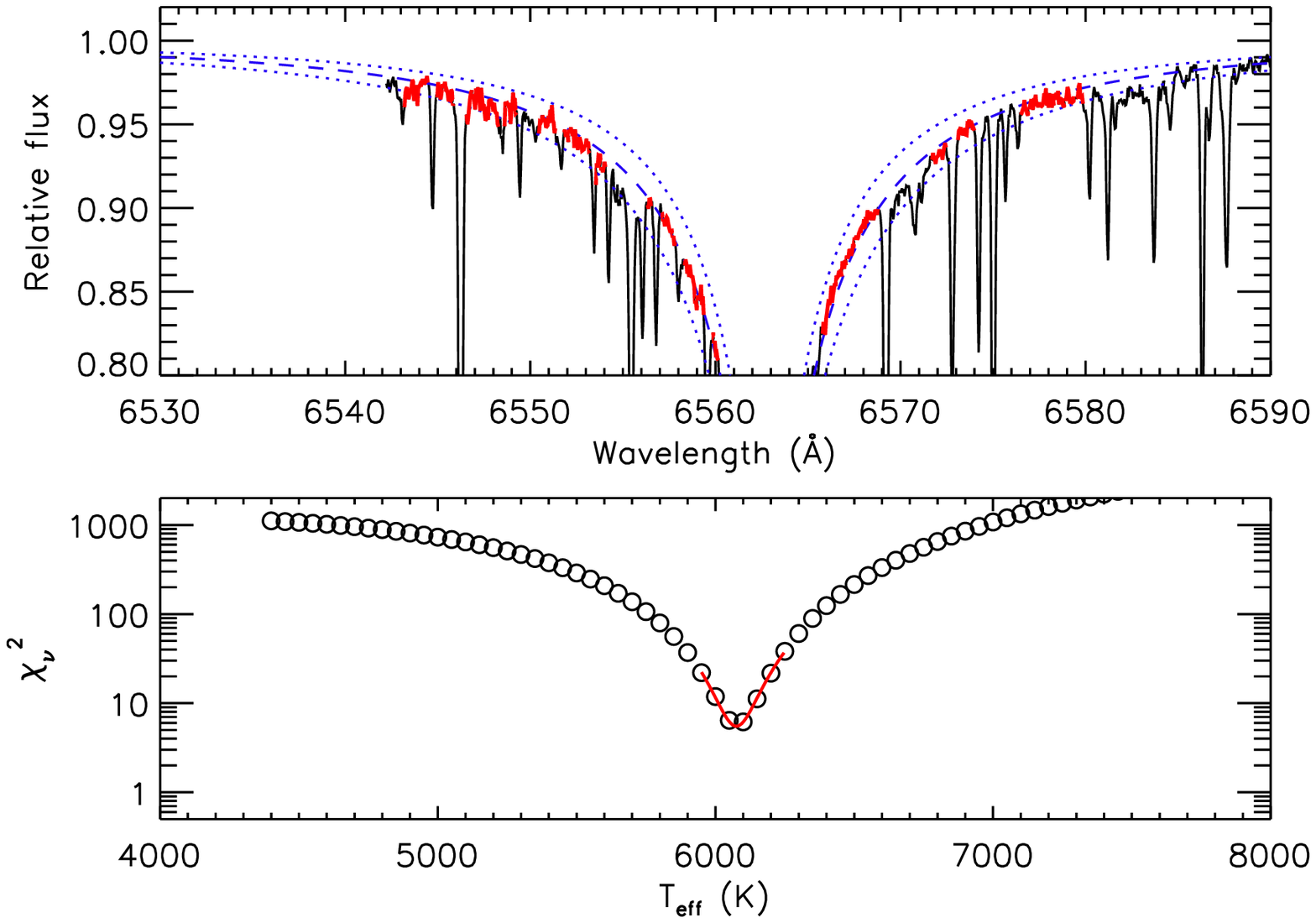}
\caption{Fits to H$\alpha$ regions of WASP-94A (left) and -94B (right)
continuum-normalized spectra to determine $T_{eff}$. The red portions
of the spectrum are the ``clean'' regions used for the $\chi^2$
minimization, where weak atomic features do not influence the fit. The
blue solid line is the best fit, and the blue dotted lines
are $\pm$200 K models. The bottom panels of both plots demonstrate the $\chi^2$ minimization procedure.}
\label{halpha}
\end{figure}

We performed several additional tests to check whether the assumed
Sun-relative parameters for WASP-94B were indeed reliable. For
$T_{eff}$, we used the [Fe/H] values from Table \ref{params} (this
work, Solar Reference) in the Casagrande et al. (2010) effective
temperature-color calibration for ($V-J$), ($V-K$), ($V-H$), and
($J-K$), with $V$ as given in NV14 with assumed errors of 0.05 mag,
and $JHK_s$ as given on \texttt{Simbad}. The average resulting $T_{eff}$
values are 6216$\pm$91 K for WASP-94A and 6096$\pm$79 K for WASP-94B,
which are consistent with the Sun-relative parameters albeit with
larger errors. As another test of precise $T_{eff}$s, we matched via
$\chi^2$ minimization the observed
H$\alpha$ Balmer lines with model fits from the theoretical grid of
H$\alpha$ lines from Barklem et al. (2002), in a process detailed in
Ram\'irez et al. (2014b) that includes a non-standard 2D-normalization
of the CCD region around the line. This analysis (see Fig. \ref{halpha}) resulted in best-fit
$T_{eff}$s of 6201$\pm$25 K for WASP-94A and 6121$\pm$30 K for
WASP-94B, after applying the necessary $+$46 K offset necessary to
make the H$\alpha$ temperatures from the solar spectra match the
nominal solar $T_{eff}$ of 5777 K (as derived in Ram\'irez et
al. 2014b). The H$\alpha$ effective temperatures are even closer to
the values obtained from the A-B analysis described above (see Table
\ref{params}, This work, WASP-94B Reference), including the
$\Delta~T_{eff}$ (80 K from H$\alpha$ versus 86 K from A-B analysis).

\begin{figure}[h]
\centering
\includegraphics[width=1\textwidth]{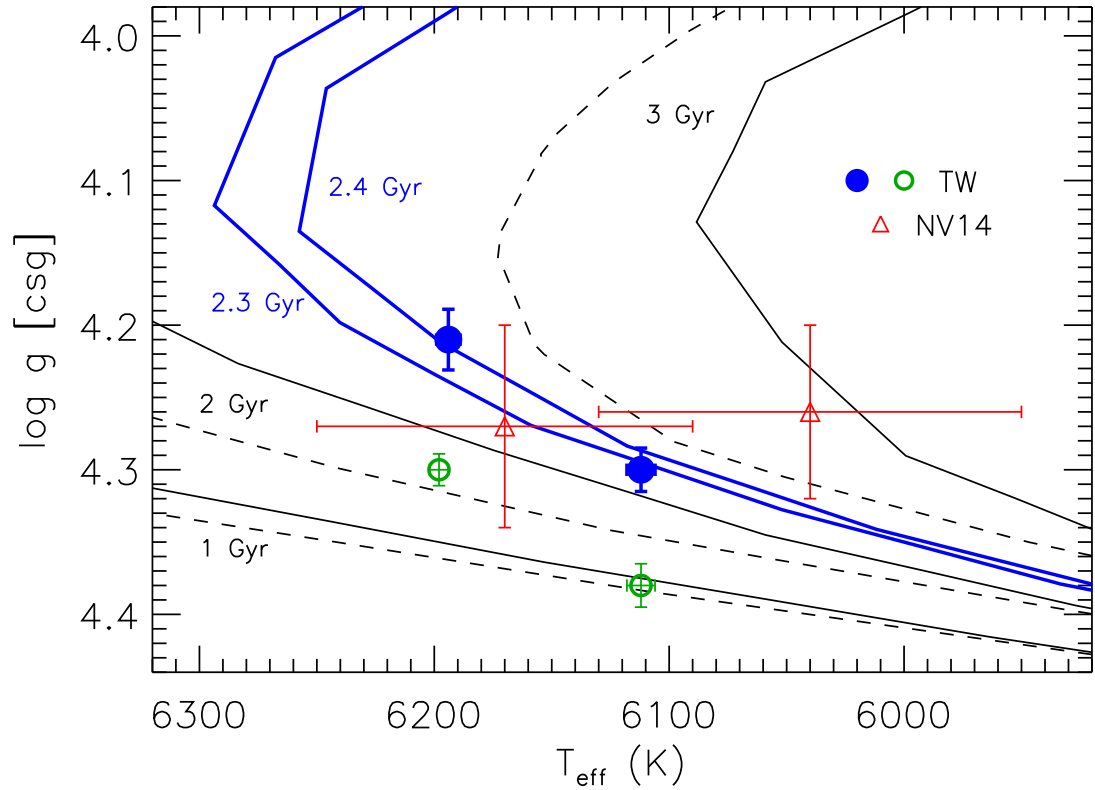}
\caption{Evolutionary state test for precise surface gravities of
  WASP-94A and -94B. The solid/dashed lines are Yonsei-Yale/Padova
  isochrones of the noted ages. The red open triangles represent WASP-94A and
  -94B with the parameters of NV14. The green open circles
  show the parameters derived in the A-B  parameter analysis, and
  listed in Table \ref{params} under This work, WASP-94B
  Reference. The blue filled circles represent the same
  $T_{eff}$ values as the green points, but with log $g$ values
  shifted by $-$0.08 dex, providing better agreement for WASP-94A and
  -94B ages. }
\label{iso}
\end{figure}

With confirmation of the $T_{eff}$ values derived in the A-B
differential analysis, the derived log $g$ values can be checked
against theoretical isochrones; if the stars are bound and coeval, their ages
should be consistent. In Figure \ref{iso}, the A-B parameters are
plotted against Yonsei-Yale and Padova isochrones, and it is clear
that the A-B log $g$ values result in different ages for the two
stars (green open circles). However, if we assume that the surface gravities
of both stars are 0.08 dex less (blue filled circles), then the stars both fall
nicely on the 2.3 or 2.4 Gyr Yonsei-Yale isochrone (solid blue
lines; both are consistent with 2.4 Gyr within errors). Thus, we fixed
the log $g$ of WASP-94B to 4.30 dex, rather than 4.38 dex as
originally determined relative to the Sun, and rederived the relative
WASP-94A stellar parameters for each set of EW measurements. This
resulted in very small changes in the other derived stellar parameters of WASP-94A
(see Table \ref{params}, This work, WASP-94B Reference, Isochrone Log
$g$). Given the lower log $g$ values are more consistent with the same
age for both stars, these are used in the subsequent elemental
abundance analysis. Lowering the log $g$ values by 0.08
(B)/0.09 (A) dex did not result in changes to the elemental abundance
ratios (detailed in the next section) outside
the original errors derived from the higher log $g$ values. 

The isochrone analysis presented here indicates that the stars have an
age of $\sim$2.5 Gyr (with an error of a few tenths of a Gyr),
significantly younger than the age reported in NV14, $\sim$4 Gyr. The
best matching age based on the Yonsei-Yale isochrones in Figure
\ref{iso} (blue solid lines) is 2.3-2.4 Gyr, whereas the best matching
age based on the Padova isochrones is more like 2.8 Gyr, still
significantly younger than the 4 Gyr age reported in NV14.

\subsection{Elemental Abundances}

With knowledge of the environment in which spectral lines form, we can
translate EW measurements of other elements into abundances via a
curve-of-growth analysis within MOOG. The EWs of lines measured in the
Sun, WASP-94A, and WASP-94B corresponding
to 23 elements, including Fe, are listed in Table \ref{lines}. Our
procedure for measuring EWs in WASP-94A and -94B included a direct
comparison in the normalized and Doppler-corrected spectra of every
line. This allowed us to choose a continuum region that was the same in
both spectra, reducing potential systematic error, and check for
differences in the lines that may affect the upper and lower limits of
the EW fit (e.g., if we want to avoid a blend). Carbon
abundances were measured from both C I and CH lines, and for Sc, Ti,
and Cr lines from both neutral and singly-ionized species were
measured (in addition to Fe). Hyperfine structure corrections were employed for V, Mn, Co,
Cu, Rb, Y, and Ba, and we applied the O triplet non-LTE corrections from
Ram\'irez et al. (2007). 
The relative abundances obtained from a line-by-line analysis, averaged
over three independent EW measurements, are listed in Table \ref{abuns},
along with errors that add in quadrature both the line-to-line scatter
($\sigma/\sqrt{n-1}$) as well as the errors propagated from each
parameter uncertainty. For species with only one line (K I, Zr II), we
adopted the largest line-to-line scatter for species with more than
three lines available, within each set of independent EW measurements
(the value of the largest line-to-line scatter was different for each of the 
measurement sets). 

The NLTE corrections to the O I triplet line around 7775 {\AA} are
uncertain, and several different groups have published
corrections. While our strictly differential approach between the two
``twin'' star should help eliminate uncertainty in the $\Delta$[O/H]
values, we checked our results by performing a synthesis analysis of
the [O I] line at 6300 {\AA} using the MOOG \texttt{synth} driver, as
outlined in Teske et al. (2014), \S 3.1.1. The synthesis fitting of
WASP-94A yielded absolute oxygen abundances (log$N$(O)) of 9.01-9.03, while fitting
of WASP-94B yielded log$N$(O) values of 8.98-9.02, confirming a small
positive $\Delta$[O/H] for (A-B) that is represented by our reported
error (0.006 dex). We also confirmed our negative $\Delta$[C/H] value
by synthesis fitting two blended C$_2$ molecular features at 5086.3 {\AA}  and
5135.6 {\AA}  via the process outlined in Teske et al. (2013), \S
2.3.1. The fitted log$N$(C) of WASP-94A ranged from 8.63-8.65, and
from 8.66-8.68 for WASP-94B, resulting in $\Delta$[C/H] (A-B) values between
-0.05 and -0.02, within our reported errors for
$\Delta$[C/H]. Importantly, we confirm that the $\Delta$[C/H] value is certainly below
the zero line. 

NV14 reported Li abundances log $N$(Li)$_{A}=$2.10$\pm$0.07 and log
$N$(Li)$_{B}$ $\leq$1.20. They propose that the WASP-94 stars have
already undergone magnetic breaking and Li depletion due to their low
log$N$(Li) and $v$sin$i$ values. We confirm the low log$N$(Li)
abundances in WASP-94AB, and a significant (A-B) difference, based on a synthesis analysis of the 6707.8 {\AA}  doublet using
the line list of Mandell, Ge \& Murray (2004). Our synthesis fitting suggests
WASP-94A's log$N$(Li)$=$2.01-2.04 and WASP-94B's log$N$(Li)$\leq$1.62,
resulting in a $\Delta$log$N$(Li)$\geq$0.39. 
Several
authors have suggested that the formation of planets could
enhance lithium depletion in host star photospheres (Gonzalez 2008, 2014;
Israelian et al. 2009; Figueira et al. 2014; Delgado Mena et al. 2015) due to
planetary migration (Castro et al. 2008), and/or star-disk
interactions (e.g., Bouvier 2008), but there
is not yet consensus due to selection biases in planet host samples
and large lithium abundance uncertainties (e.g., Ghezzi et al. 2010;
Baumann et al. 2010; Ram\'irez et
al. 2012). Both WASP-94 stars border the ``lithium desert'' proposed
by Ram\'irez et al. (2012), a region in $T_{eff}$ versus log$N$(Li)
space devoid of stars. Indeed, there are no exoplanet host stars included in Ram\'irez et al. (2012) with as low
log$N$(Li) and as high $T_{eff}$ and [Fe/H] as WASP-94B (see their
Figure 9), making WASP-94B a slightly unusual case. The star's relatively low log$N$(Li) suggests
it may have experienced some short-lived surface destruction of
lithium. Here we do not focus on the study of lithium, but the
difference between WASP-94A and -94B, a pair of hot ``twin'' stars
both hosting close-in giant planets, could prove to be a powerful tool
for better understanding whether there is a connection between lithium
depletion in stars and the presence of planets.

\section{Discussion}
\subsection{Results: $\Delta$[X/H] vs. $T_c$}
\begin{figure}[h]
\centering
\includegraphics[width=1.\textwidth]{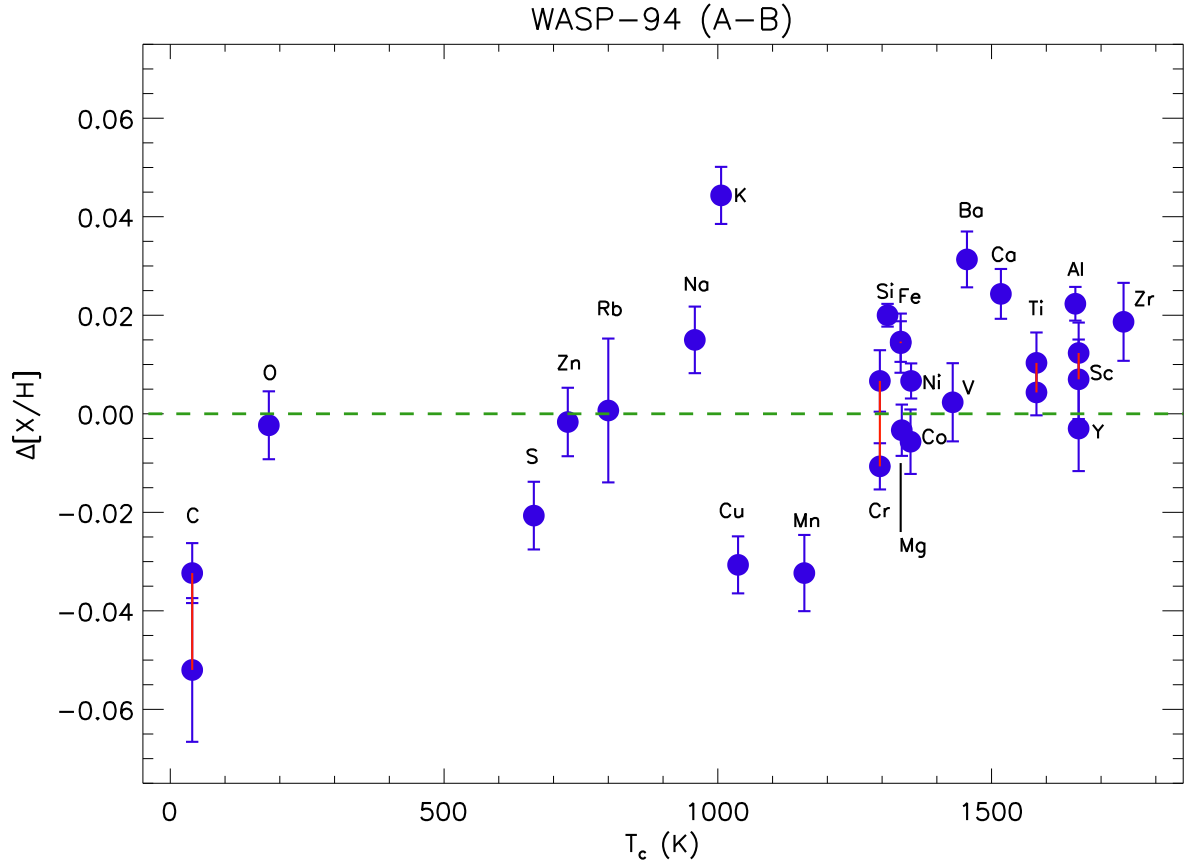}
\caption{The $\Delta$(A-B) relative
  abundances versus T$_c$ (Lodders 2003), calculated using the derived
  (A-B) stellar parameters with log $g$ of B fixed to 4.3 dex (see
  last row of Table~\ref{params}). This plot shows the abundances before removing
  absorption lines with EW$>$100 m{\AA}; Figure \ref{key2} shows the abundances 
  after removing these lines. The only significant difference
  between the abundances in the two plots is the $\Delta$[Na/H] value, which is lower in
Figure \ref{key2}.}
\label{key1}
\end{figure}

In Figure \ref{key1}, the abundance differences between the two
stars, $\Delta$[X/H], are plotted against the 50\% condensation
temperatures ($T_c$) from Lodders (2003) for solar composition
gas. There is a small but significant (see below) enhancement of high-$T_c$ elements
($\geq$1200 K) in WASP-94A relative to -94B. Lodders (2003) refers to
phosphorous at $T_c=1229$ K as ``moderately volatile'', sulfur at
$T_c=664$ K as ``volatile'', and iron, magnesium, and silicon
($T_c=1334, 1529, 1397$ K, resp.) as ``common''. As in other studies
looking for correlations between abundance differences and
condensation temperature, the $T_c$ values we
use here are for a solar-composition gas. However, $T_c$ values will
depend on the composition of the gas, which in the case of WASP-94AB
is super-solar. Detailed chemical equilibrium condensation sequence calculations
such as those in Bond et al. (2010) are beyond the scope of this work, but the sequences calculated in
that work for gas of similar composition to WASP-94AB indicate $T_c$
values do not increase more than a few tens of K. Even in
the high-metallicity condensation sequences where $T_c$ for C and O
increase by up to about 1000 K, this would not affect our overall
results because our ``break'' between volatiles and refractories is
likely higher than $\sim$1000 K (see below). Most of the lower
$T_c$ volatile or moderately volatile elements are depleted in
WASP-94A, save sodium (Na) and potassium (K). The latter is based on
one saturated line that suffers from a large NLTE effect ($\sim$0.3
dex; e.g., de La Reza and M{\"u}ller 1975; Zhang et al. 2006); this
abundance may thus not be very accurate.
In an attempt to
ameliorate the high $\Delta$[Na/H] value, as well as check for
systematic error, the elemental abundance derivation was
repeated across all elements, removing any potentially-saturated lines with EW values
$>$100 m{\AA}. The resulting values are shown in 
Figure \ref{key2}, with K and O unchanged, as their abundances
are based on only strong lines. None of the new abundance derivations
that exclude strong
lines change within the errors, \textit{except} for Na, which
is exactly zero when two 
strong lines are removed from the
average; removing these lines also reduces the line-to-line scatter, slightly
decreasing the $\Delta$[Na/H] error. 

\subsubsection{Linear 1- and 2-Component Fits}
\begin{figure}[h]
\centering
\subfigure{\includegraphics[width=.75\textwidth]{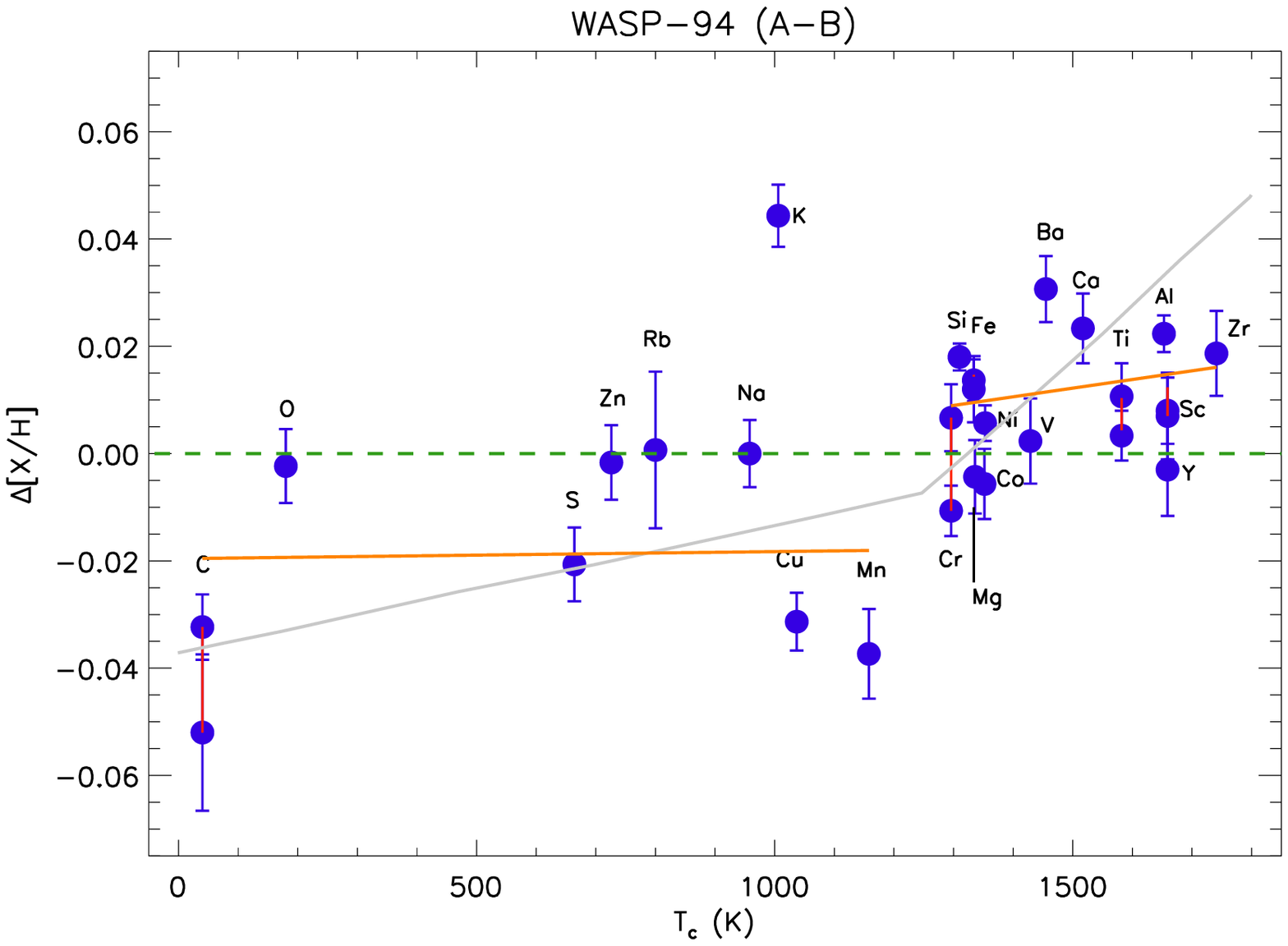}}
\quad
\subfigure{\includegraphics[width=.75\textwidth]{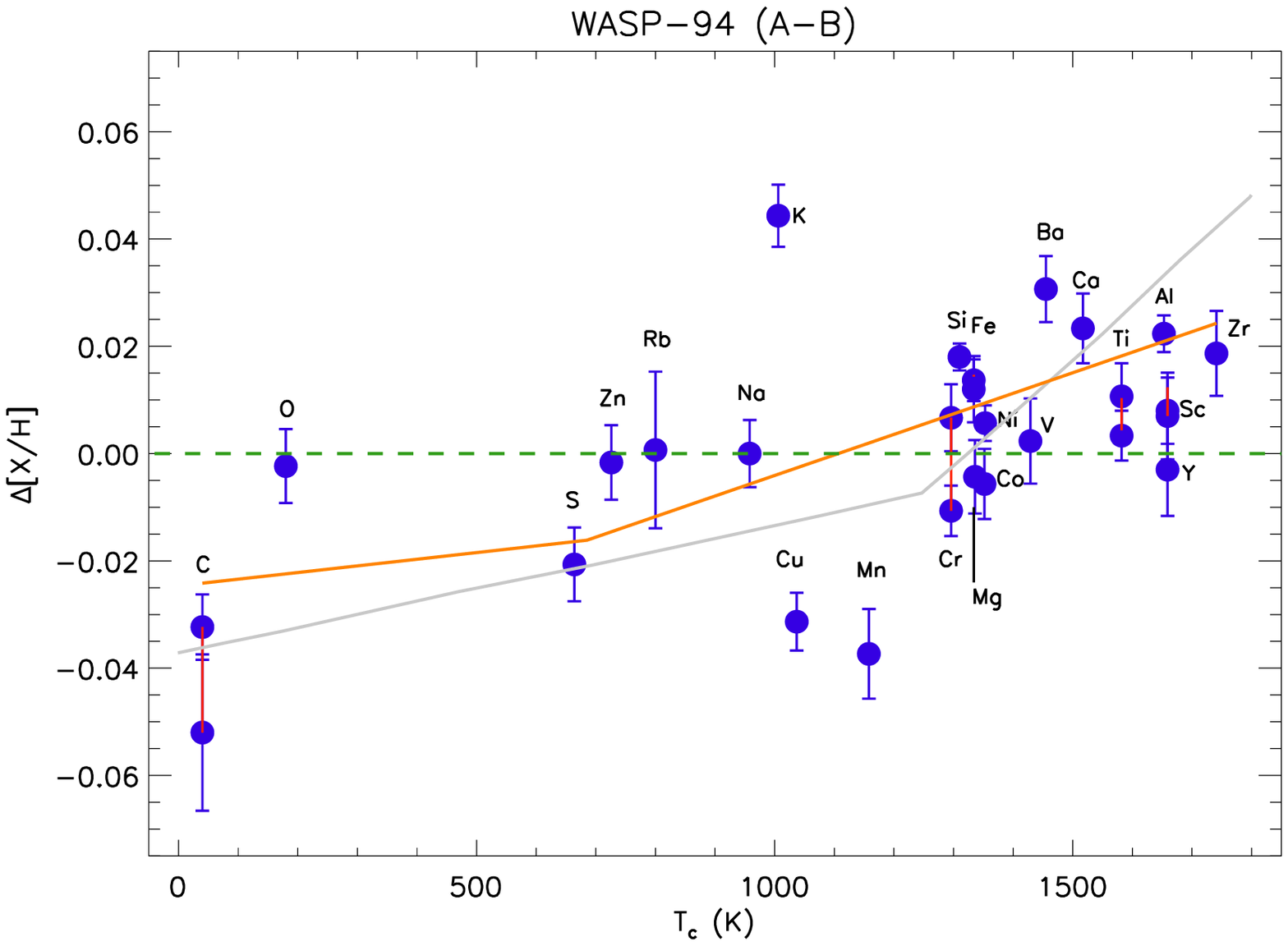}}
\caption{The $\Delta$(A-B) relative
  abundances versus T$_c$ (Lodders 2003), calculated using the derived
  (A-B) stellar parameters with log $g$ of B fixed to 4.3 dex (see
  last row of Table~\ref{params}). These plots show the abundances after removing
  absorption lines with EW$>$100 m{\AA}; O and K have only strong lines so are
  not removed. The only significant difference
  between the abundances in these plots and Figure \ref{key1} is the
  $\Delta$[Na/H] value, which is lower here. Also shown in orange are the best fits to the abundances,
not including K, as discussed in \S 4.1.1, without (top) and with (bottom)
a smooth $T_c$ break point. The grey lines shows the solar twin
trend from Mel{\'e}ndez et al. (2009) with a constant 0.015 dex added
to account for the [Fe/H] difference between A and B.}
\label{key2}
\end{figure}

Examining the new abundances in Figure \ref{key2} and excluding
K, there appears to be a natural break point around $T_c=1200$ K,
between Mn and Cr, although with moderate scatter ($\sigma$ of all
$\Delta$[X/H]$=$0.02 dex). The $\chi^2$ value\footnote{$\chi^2 =
  \sum$[(observed-predicted)$^{2}$/error$_{\rm{observed}}$]} of a simple
zero-slope, zero-intercept fit (the dashed green line in Figure \ref{key2}) is 1.48. A weighted\footnote{by 1/$\Delta[X/H]_{err}^2$} linear fit to the data with a forced-zero
slope but a free intercept results in an intercept$=0.0055 \pm 0.0010$ dex, and a
$\chi^2$ value of 1.50, slightly worse than a flat line with intercept$=0$. A
linear fit to the data with an unconstrained slope results in an
intercept of -0.0302$\pm$0.0037 dex and a slope of
2.7723$\pm$0.2795$\times$10$^{-5}$ dex/K, with a $\chi^2$ value of
0.86 and a mean scatter around the fit of -0.0035 dex. The data thus support a non-zero slope at the $\sim$10$\sigma$
level. 

To test whether a two-component linear function is favored (results in a
lower $\chi^2$ value), we first repeat a linear fit to
all the points (except $\Delta$[K/H]), removing the
highest-$T_c$ point from each iteration to find the number of points whose linear fit
results in the lowest slope value. The slope minimum occurs at
$T_c=$1158 (Mn). This minimum is then used as $P[2]$ in a custom
fit, executed with the IDL \texttt{MPFITFUN} routine (Markwardt 2009), with the functional form $[P[0]+P[1]\times
X[0:P[2]],P[3]+P[4]\times X[P[2]+1,*]]$, where $P[0], P[1], P[3]$ and
$P[4]$ are allowed to vary, $X$ corresponds to $T_c$, and $*$ is the
last entry in $X$. The
resulting best-fit parameters are $P[0]=$-0.0197$\pm$0.005 dex,
$P[1]=$1.463$\pm$5.996$\times 10^{-6}$ dex/K,
$P[3]=$-0.0120$\pm$0.0114dex and
$P[4]=$1.615$\pm$0.795$\times 10^{-5}$ dex/K. This two component
linear fit has a $\chi^2$ value of 0.71, with a volatile element ($T_c
< 1200$ K) slope
consistent with zero and a refractory element ($T_c >1200$ K) slope
inconsistent with zero at the 2$\sigma$ level. This two component fit
is shown in green in the top panel of Figure \ref{key2}. The mean
scatter around this two-component fit is -0.0023 dex, and using $T_c$=1200 K as the divider,
we find weighted averages (and weighted standard deviations) of
$\Delta$[X/H]=-0.019$\pm$-0.006 for volatiles (i.e., a deficiency detected at the 3$\sigma$ level) and $\Delta$[X/H]=$+$0.011$\pm$0.002 for refractories (i.e., an enhancement detected at the 5$\sigma$ level).

As a second test of the $T_c$ break point, we performed two separate linear fits
to the measurements, and varied the $T_c$-break temperature (where one fit
ended and the other began) between 500 K and
1500 K. For each $T_c$-break temperature, we calculated the difference
between the end of the low $T_c$ fit and the beginning of the high
$T_c$ fit to find the value that resulted in a smooth transition between the two linear
fits (zero difference between the end/start points). This $T_c$-break
temperature is 685 K, much cooler than the 1200 K value found
above. In this fit, the lower $T_c$ component has a slope
1.241$\pm$4.756$\times$10$^{-5}$ dex/K and an intercept of
-0.0247$\pm$0.0167 dex, while the higher $T_c$ component slope of
3.829$\pm$1.233$\times$10$^{-5}$ dex/K and an intercept of
-0.0424$\pm$0.0175 dex. Using $T_c$=685 K as the divider, we find weighted averages (and weighted standard deviations) of
$\Delta$[X/H]=-0.022$\pm$-0.008 for volatiles (i.e., a deficiency detected at
the 2$\sigma$ level) and $\Delta$[X/H]=$+$0.008$\pm$0.003 for refractories
(i.e., an enhancement detected at the 2$\sigma$ level). However, the resulting fit shown in the bottom panel of
Figure \ref{key2} has a $\chi^2$ of 0.936 and a mean
scatter around the fit of -0.0053 dex, a slightly worse fit than the
2-component fit described above with a $T_c$-break point of 1200 K.

\subsubsection{Nature of Scatter}

\begin{figure}[h]
\centering
\includegraphics[width=1.\textwidth]{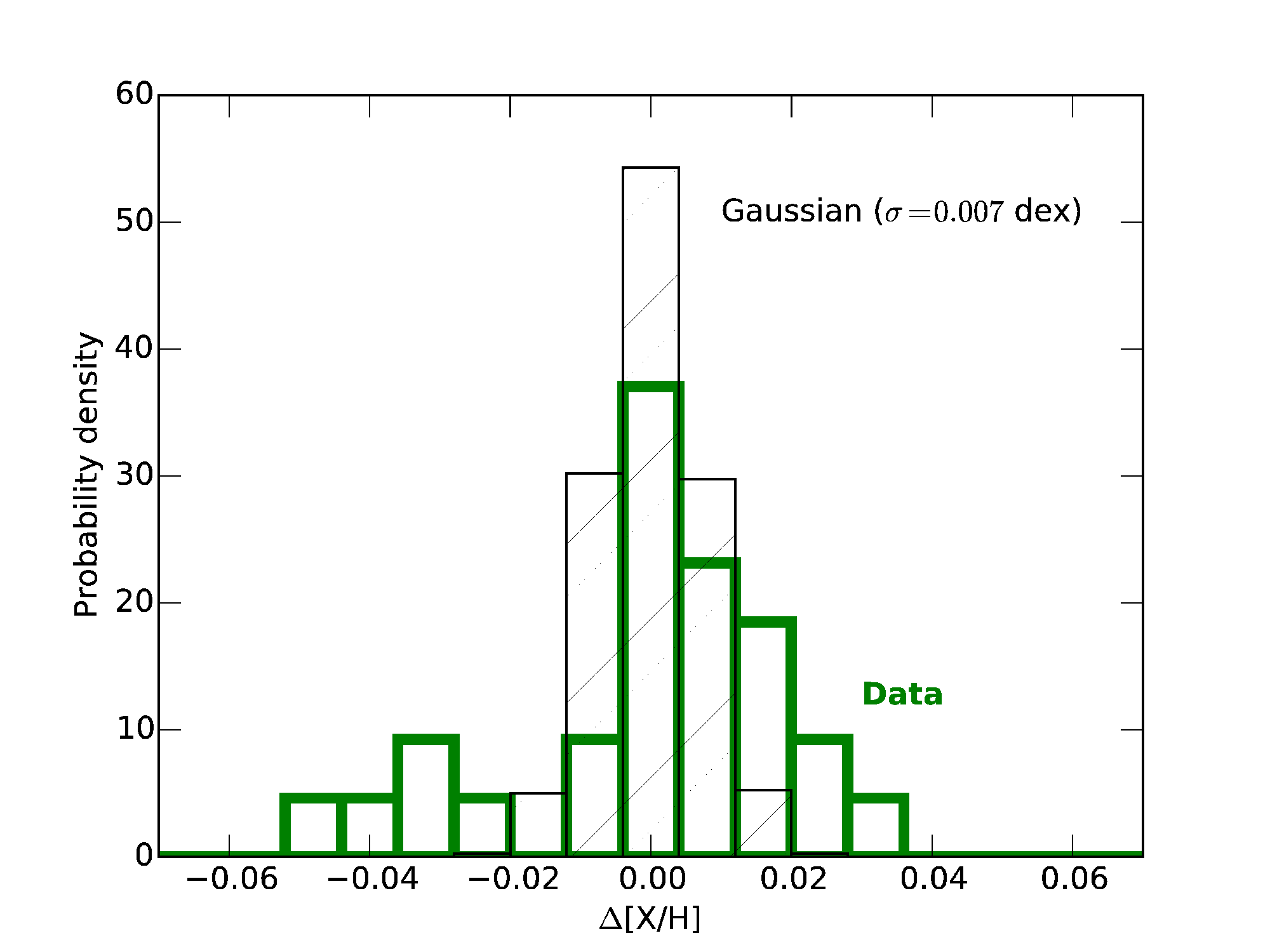}
\caption{A histogram of normalized probability density of
  $\Delta$[X/H], comparing a sample ``S'' of 100,000 values randomly
  selected from a Gaussian distribution with $\sigma$=0.007 dex (black
  dashed bars) and our actual data (green
open bars). It is clear that our data do not follow a normal
distribution, and both a two-sample KS test and an AD test indicate
the two samples are significantly different (see \S 4.1.2)}
\label{stats}
\end{figure}

We ran standard statistical tests to address the question of
whether the scatter seen in Figure \ref{key2} is consistent with random
observational noise. These tests are independent of the $T_c$
correlation and only examine the level of scatter and its
normality. For its use in some of these tests, we created a sample
``S'' of 100,000 $\Delta$[X/H] values randomly selected from a
Gaussian distribution of $\sigma$=0.007 dex, which is the average
error bar of our $\Delta$[X/H] values. A two-sample Kolmogorov-Smirnov
(KS) test gives a $p$-value of 0.06 when comparing the actual data to
the pure noise sample $S$. An Anderson-Darling (AD) test on the same
two samples gives a significance level of 2.5$\times$10$^{-5}$. We
repeated these tests assuming that our error bars had been
underestimated and redefined the $S$ sample assuming Gaussian noise of
$\sigma$=0.010 dex instead. In this case, we find a much larger KS
$p$-value of $\sim$0.4, which would imply that our data could be
compatible with no abundance difference within error, but the
significance level from the AD test remains low, at 0.03. The AD
test is known to be superior to the KS test for a variety of reasons,
in particular its higher sensitivity to the edges of the distributions
(see, e.g., Feigelson \& Babu 2012, their Sect. 5.3.1). We also
employed the AD and Shapiro-Wilk (SW) tests to quantify the
``normality'' of our $\Delta$[X/H] data set (these tests do not
require an estimate of the sigma value a priori). In both cases, the
statistic derived corresponds to a significance level below 6\%. In
other words, the probability that our $\Delta$[X/H] values correspond
to a Gaussian distribution centered at zero is low. Despite having
relatively few data points, the non-normality of our abudance
differences can be clearly seen in Figure \ref{stats}. Thus, these statistical tests suggest that there are in fact true
abundance differences between the stars in the WASP-94 system.

As noted in \S 1.1, WASP-94 differs from the other known twin binary
systems in which both stars host planets in several important ways. The trend with $\Delta$[X/H]
and $T_c$ observed here -- more
refractory elements in WASP-94A than B, but fewer volatile elements --
also differs from what is observed in the other known twin
binary systems, including those in which only one star is known to
host a planet. In every other published case, either no significant
abundance differences are reported, or there is some degree of enhancement in
\textit{all} elements, including those with low $T_c$. The pattern
observed here in WASP-94AB appears qualitatively similar to that
reported by some authors in the 16 CygAB system (Ram\'irez et
al.\,2011; Tucci Maia et al. 2014) and reported in the XO-2AB
system (Teske et al. 2015; Ram\'irez et al.\,2015; Biazzo et
al. 2015) -- volatiles have a $\sim$flat slope, while refractories
have a positive slope -- but in WASP-94A the volatiles are depleted,
not enhanced like in 16 Cyg A and XO-2N. Additionally, the differences we find here
between WASP-94AB are smaller than those seen in 16 CygAB and XO-2AB,
causing the appearance of our Figure \ref{key2} to be ``noisier'' due
to the fact that it is ``zoomed in'' more than the previous cases (not
because our errors are larger). Note, for example, that the
enhancement of refractory element abundances compared to volatiles in
solar twins relative to the Sun is about 2-3 times larger than that
seen in WASP-94 A minus B, as shown by the thin gray line in Figure
\ref{key2}.

\subsection{Explanation of Observed Trend}
What could account for the observed trend in $\Delta$[X/H] abundances?  Ram\'irez et
al. (2015) explained the volatile depletion in XO-2S as potentially ``more'' gas
giant planet formation around it versus XO-2N, leading to a relatively larger amount of
volatile elements in -2N. Similarly, Tucci Maia et al. (2014) suggested the volatile
depletion in 16 Cyg B relative to A as was due to the envelope of the giant
planet around 16 Cyg B, and the refractory depletion was a signature
of the rocky core of 16 Cyg Bb. Examining Figure \ref{key2}, the
volatile depletion in A could be explained by (more) giant planet
formation causing a decrease in its overall metallicity, and thus
volatile abundances, relative to B, and (more) rocky planet formation around B
causing a decrease in its refractory abundances relative to A. 

Alternatively, the observed abundance differences could be explained by the difference in metal content
of WASP-94Ab and -94Bb, and not any additional planets. Miller \&
Fortney (2011) compared interior models of 14 cool transiting giant
planets to their host stars' [Fe/H] values, and found that the overall
heavy element mass for planets ($M_z$) increases with planet mass, but that the percentage of metals in
the planet versus the percentage of metals in the star ($Z_{pl}/Z_{star}$) decreases with planet mass, consistent
with Solar System giant planets. If WASP-94Ab was larger, and thus had
a smaller refractory-to-volatile abundance ratio than WASP-94Bb, then
WASP-94A might contain less volatile but more refractory material ``left
behind'' after the formation if its giant planet. 

However, both of these explanations are unsupported by the
planets actually observed. The planet around WASP-94B
is more massive, so -94B should be the star to show the lower overall
metallicity and volatile abundances, or the higher
refractory-to-volatile ratio of ``left over'' material in the star. We can estimate whether the mass
difference between WASP-94Ab and -94Bb can even explain the observed
abundance differences using the corrected formula from Ram\'irez et
al. (2011):

\begin{centering}
\begin{equation}
\Delta[M/H] = \mathrm{log}\bigg[\frac{(Z/X)_{cz}M_{cz} + (Z/X)_PM_p}{Z/X_{cz}(M_{cz}+M_p)}\bigg]
\end{equation}
\end{centering}

\noindent with the convection zone mass $M_{cz}$ estimated as 0.0032
M$_{\odot}$ from the relation in Pinsonneault
et al. (2001), $(Z/X)_p$ estimated as 0.1, $M_p$ as the mass
difference between Ab and Bb (0.166 M$_J$, using Bb's $Msini$), and
$(Z/X)_{cz}$ estimated by scaling Asplund et al. (2009)'s
$(Z/X)_{\odot}=0.134$ to WASP-94A's [Fe/H] (0.32 dex). The required
$\Delta$[M/H] for 0.166 M$_J$ of material ``added to'' or ``missing
from'' WASP-94A is $\sim$0.05 dex, much higher than the
volatile/refractory abundance differences of 0.02/0.01
dex. A similar excercise for WASP-94B still results in $\sim$0.03 dex
of material, which is more than is ``missing'' or ``added'' to
WASP-94B. Furthermore, if as suggested as one source of the refractory enhancement in XO-2N (Teske et
al. 2015; Ram\'irez et al. 2015; Biazzo et al. 2015), the refractory
enhancement seen in this system is due to the migration of a giant
planet pushing/dragging rocky material on to the star, why is this accretion
signature seen in the star (A) with the \textit{least} massive planet,
which would presumably have a smaller gravitational effect on rocky material in the
disk? 

One can reverse the accretion equation, and instead solve for the mass
of material that could explain the missing 0.02 dex of volatile/excess
0.01 dex of refractory material in WASP-94A (or the excess
0.02/missing 0.01 dex in -94B). In WASP-94A, which likely has a
somewhat smaller convection zone than the cooler -94B, a deficit of
0.02 dex of $(Z/X)_p=0.1$ material requires $\sim$0.063 M$_J$ of
material to be ``missing''. In WASP-94B, an enhancement of 0.02 dex of $(Z/X)_p=0.1$
material requires $\sim$0.098 M$_J$ of material to be ``added''. These masses are
likely too small to be single bodies responsible for dynamical
effects, e.g., a giant planet affecting the orbits and/or migration of
either WASP-94Ab or -94Bb (Chatterjee et al. 2008; Ford \& Rasio 2008). However, such masses could correspond to
super-Earth or Neptune-sized planets (or amounts of material). This material could have been
scattered inward (increasing the stellar abundance) or outward
(decreasing the stellar abundance) by planet-planet interactions that may
have caused the orbit misalignment and retrograde orbit of
WASP-94Ab (e.g., Bromley \& Kenyon 2011; Bromley \& Kenyon 2014). Alternatively, these bodies could have been accreted onto
the star during the inward migration of WASP-94Bb  (Raymond et al. 2011; Fogg \& Nelson 2005), which has near-zero 
$e$ (0.13$\pm$0.20, NV14). Figure 1 from Kaib et al. (2013)
suggests that a binary like WASP-94 has a $\sim$0.5 instability
fraction, and a rather large critical pericenter ($\sim$200 AU),
meaning that if the binary separation evolved at all during the short
lifetime of the stars, an instability would be more likely.

The above estimations assume a bulk $(Z/X)_p=0.1$ for planetary mass
material, based on what is known about the giant planets in the Solar
System (Guillot 2005; Fortney \& Nettelmann 2010) and other giant
exoplanets (Miller \& Fortney 2011). However, if this value were
smaller or larger, the effect on the stellar abundances would
correspondingly decrease or increase. The value of $(Z/X)_p$ is not
constant across Solar System planets, nor is it independent from the
size of the envelope versus the core of the planet (e.g., Fortney \&
Nettelmann 2010). The proportion of volatile to refractories
depends on the formation location, mass, and interior differentiation
of planets; our model above is only a first-order estimation and
future studies should more thoroughly test the impact of differentiated
planet formation on stellar abundances. 

The above calculations also assume a current $M_{cz}$ based on the
mass of the star, but $M_{cz}$ is known to vary greatly across stellar
lifetimes (Hayashi 1981; D'Antona \& Mazzitelli 1994; Serenelli et
al. 2011). Ram\'irez et
al. (2011) explored how varying both $M_{cz}$ and $(Z/X)_p$ changed
the amount of material required to account for the abundance
difference between 16 Cyg A and B (see their Figure 12), based on both
standard stellar models (Serenelli et al. 2011) and non-standard
models of episodic accretion (Baraffe \& Chabrier 2010). At earlier times when the stellar
convection zone is a significant fraction ($\gtrsim$10\%) of the mass of the star,
a larger or more metal-rich amount of accreted material is required to cause
the same stellar abundance change. For instance, a $M_{cz}=0.01
M_{\odot}$ for WASP-94A would require $\sim$0.20 $M_J$ of material to
cause a 0.02 dex difference, over 3$\times$ as much material than the
smaller (current) $M_{cz}=0.0032 M_{\odot}$ case. This might be one way
to help explain a mass difference of $\geq$0.166
$M_J$, as observed between WASP-94Ab and -94Bb, but again the trend is
opposite of what is expected (A is depleted by $\sim$0.20 $M_J$, not
B), and it still leaves the refractory abundance enhancement in WASP-94A unresolved.

Perhaps what is most intriguing is why two ``twin'' similar stars formed
apparently different planets. The planet around WASP-94A is less
massive and misaligned with the stellar orbital axis, and likely
retrograde, both signatures of dynamical interactions in the history
of the planet's formation and evolution. Why does WASP-94B appear to be circularized? The $T_{eff}$ of both stars put them near the $T_{eff}$ border that Winn et al. (2010) and
Schlaufman (2010) suggest separates systems in which tidal dissipation
damps planetary obliquities within a few Gyr and systems in which
dissipation is ineffective (stars hotter than 6250 K). Winn et
al. (2010) explain this break point as where the convection zone mass of stars becomes negligible. Based on the relation in Pinsonneault
et al. (2001), WASP-94A's $M_{cz}$ is between
$\sim$0.003 and 0.004 M$_{\odot}$, while WASP-94B's is between $\sim$0.005
and 0.006 M$_{\odot}$. Perhaps the 2$\times$ larger convection zone of
WASP-94B versus -94A was enough to better facilitate tidal circularization of its giant planet. 

Another possibility is that the abundance differences are due to or
influenced by the orbital dynamics of the system. There may be a very stellar distant companion to
WASP-94A or -94B, which would not have formed out of the
protoplanetary disk and thus be undetectable in the abundance
signatures, nor detectable in the limited RV coverage ($\sim$2 years). We are
conducting high-contrast imaging observations to search for any such
wide, massive perturbers, which may or may not be bound to either star, that could provide clues to the dynamical history of this system.

It may be that the small differences that have been observed
in the high-precision 
abundances of planet-host stars are not related to planet
formation. \"{O}nehag et al. (2011) suggest that the Sun's peculiar
abundance trend with $T_c$ is not due to small planet formation, but removal
of refractory dust material early on in the Sun's life. They applied a strictly differential
(versus the Sun)
approach to M67-1194, a solar twin in the M67 open cluster, to
minimize systematic errors and determined not only that its parameters
are indistinguishable from the Sun (except for [Fe/H]= 0.023$\pm$0.015
dex), but also that its abundance pattern very closely resembles the
Sun, unlike the solar twins in Mel\'endez et al. (2009). The authors
suggest that M67-1194 and the Sun may be from the same cluster, or at
least that the Sun was born in a cluster similar to M67, and that both
stars were affected during
their fully-convective phase by dust cleansing by luminous stars in
the same cluster. In \"{O}nehag, Gustafsson, \& Korn (2014), the authors
expand their sample to 14 M67 stars and find a similar agreement
between the stars' abundances and the Sun's, which they use as further
support for their hypothesis of dust cleansing in both protostellar
clouds that formed the M67 stars and the Sun (which are perhaps the
same). (We note, however, that Pichardo et al. (2012) present strong
dynamical arguments that reject M67 as a parent cluster of the Sun.) Adibekyan et al. (2014) explored a large sample of
solar-like stars and found possible correlations between the
$\Delta$[X/H] vs. $T_c$ slopes and stellar ages, and between the
slopes and mean galactocentric distance. The authors thus suggest that
differences in $\Delta$[X/H] vs. $T_c$ slopes may not be related to
planet formation at all, but instead to the age and galactic birth
place of a given star (see, however, Spina et al. 2015). Gaidos (2015) pointed out that the trend of
increasing $\Delta$[X/H] with $T_c$ seen in Mel\'endez et al. (2009) is the
opposite of the pattern of element depletion in interstellar medium
(ISM) gas (Yin 2005), and suggested that it may simply be a signature of
gas-dust segregation and the particular composition of the dust in the
protoplanetary disk. Unfortunately, none of the above explanations 
can naturally explain the differences reported here between WASP-94A
and B if the assumption that they formed from the same gas cloud and
are coeval holds. 

Finally, the $\Delta$[X/H] values we
measure may be due to the different rotation or granulation in
WASP-94A versus -94B. NV14 measured $v$sin$i$ of 4.2 km~s$^{-1}$ for
WASP-94A and $<$1.5 km~s$^{-1}$ for WASP-94B; while rotational
broadening will not affect the total absorption (measured EW) of an individual line, rotational broadening can make blends
harder to exclude. The iron line list used here, from Ram\'irez et
al. (2014), was constructed specifically to exclude blended lines, but
they are not always avoidable for other elements if there are
only a few observable lines. In such cases (of
which there are only a few; most elements measured have more than a few lines),
we are careful to choose the same continuum and line boundaries in
both stars such that the blended feature does not influence the EW
fit. NV14 also used the cross-correlation function (CCF)
bisector spans, measured as part of their radial velocity analysis, to
confirm the planetary nature of WASP-94Ab and -94Bb, and found a
$\sim$25 m~s$^{-1}$ difference in the average CCF bisector values
between the two stars (see their Figures 2 \& 4). Line bisectors --
the midpoints of the horizontal segments across the wings of a
spectral line -- are a diagnostic of granulation in stellar
atmospheres. However, NV14 do not measure the bisector of spectral
lines, but the CCF. 
Also, the 25 m~s$^{-1}$ (A-B) difference is an
order of magnitude smaller than the absolute bisector span measured in
T$_{eff} \sim$6000 K stars (e.g., Ram\'irez et al. 2008), which is
already below the resolution of our spectra where the lines have
been smoothed to be almost Gaussian. Thus, the CCF bisector span
differences between WASP-94A and -94B are likely not meaningful to our
differential abundance analysis. 

\section{Conclusions}

Here we report differences in the abundances of two hot Jupiter
hosting F-type stars that are in a wide ($\sim$2700 AU) separation
binary. The stars are ``twins'', with an (A-B) $\Delta$T$_{eff}=$82$\pm$7 K,
$\Delta$log $g=$-0.08$\pm$0.019 dex, $\Delta$[Fe/H]$=$0.014$\pm$0.006 dex, and
$\Delta \xi=$0.12$\pm$0.014 km~s$^{-1}$. The abundance analysis presented here
is strictly differential, allowing a minimization of systematic
errors, and includes the average across three independent equivalent width measurement
sets, further reducing observational errors. One tangible result of this work is the reduction in age of the host
stars by at least 1 Gyr. As a result of our
careful stellar parameter determination, we could pinpoint the
isochrone on which both WASP-94A and -94B fell, based on their
respective $T_{eff}$s and log $g$s. This results in an age between 2.3
and 2.8 Gyr (not including measurement error of a few tenths of a
Gyr), depending on the isochrone used, which is significantly younger than the $\sim$4 Gyr age reported in NV14.

We find evidence of a
non-zero slope trend between $\Delta$[X/H] vs. $T_c$ at the 10$\sigma$
level, with volatile element depletion (by $\sim$-0.02 dex, on average) and
refractory element enhancement (by $\sim$0.01 dex, on average) in
WASP-94A. This differs from all other stellar abundance studies of binary systems in
which one or both stars host planets, and does not seem to match any previous
explanations of such trends, due to the mass and orbital period
differences between WASP-94Ab and -94Bb. While there is not an obvious
explanation for these abundance differences right now, the precision achieved here, at the
$\sim$0.005 dex level, allows us to examine differences that may have
been effectively washed out in previous studies of binary stars, which
suggested that anywhere from $\lesssim$0.015 dex to $\lesssim$0.03 dex
differences were to be expected.

Whatever the cause, the abundance differences detected here for
WASP-94AB, along with those seen in 16 CygAB and XO-2AB, challenge the
long-standing assumption that stars in binary systems must share the
same chemical composition at all times. The stars may have formed from a single
composition cloud, but processes that occurred after their birth have
certainly altered their surface compositions at the few percent
level. A larger sample of ``twin'' systems
with constraints on planet presence is necessary to better understand
how curious the WASP-94AB system is or not.

\acknowledgements
The authors wish to thank David Lambert for helpful
comments that improved the quality of the paper. J.K.T. wishes to
thank Stephen Shectman, John Chambers, Wladimir Lyra, Larry
Nittler, Kevin Schlaufman, and Cayman Unterborn for useful discussion that
added to the interpretation of results in this work. We thank the referee for their comments and edits that improved the paper. 

{\it Facilities:} \facility{Magellan:Clay (MIKE)}


\begin{deluxetable}{lcccccccccc}
\rotate
\tablecolumns{11}
\tablewidth{0pc}
\tabletypesize{\scriptsize}
\tablecaption{Measured Lines \& Equivalent Widths \label{lines}}
\tablehead{ \colhead{Ion} & \colhead{$\lambda$} & \colhead{$\chi$} &
  \colhead{log $gf$} & \colhead{EW$_{\odot}$} &
 \colhead{\underline{WASP-94A}}
 &\colhead{\underline{WASP-94A}}&\colhead{\underline{WASP-94A}}&
 \colhead{\underline{WASP-94B}}& \colhead{\underline{WASP-94B}} &\colhead{\underline{WASP-94B}}\\
 \colhead{ } & \colhead{({\AA})} & \colhead{(eV)} & \colhead{(dex)} &  \colhead{(m{\AA})} & \colhead{EW$_{IR}$ (m{\AA})} &
  \colhead{EW$_{SK}$ (m{\AA})} & \colhead{EW$_{JT}$ (m{\AA})} & \colhead{EW$_{IR}$ (m{\AA})} &
  \colhead{EW$_{SK}$ (m{\AA})} & \colhead{EW$_{JT}$ (m{\AA})} }
\startdata
Fe I &	4389.245	&0.052&	-4.583&	73.2	&74.2&	73.4	&74.1&
75.6	&74.8  & 75.9\\
Fe I&	4445.471	&0.087&	-5.441&	40.4	&35.8&	38.2&	36.4&	38.6	&40.4&39.7\\
Fe I&	4602.001	&1.608&	-3.154&	72.3	&78.7&	79.0&	79.4	&78.8&	79.6&80.2\\
Fe I&	4690.14	&3.69&	-1.61&	59.5	&66.9&	65.7&	66.8	&66.7&	66.1&67.6\\
Fe I&	4788.76	&3.24&	-1.73&	67.7	&74.8&	74.4	&73.1&	74.3	&74.5&72.6\\
Fe I&	4799.41	&3.64&	-2.13&	36.0&37.7&	42.1	&41.0&	40.0	&42.3&41.4\\
Fe I&	4808.15	&3.25&	-2.69&	27.6	&28.9&	28.6	&29.3&	31.0	&31.3&30.9\\
Fe I&	4950.1	&3.42&	-1.56&	74.6&80.5&	80.8	&89.3&	81.0	&81.3&88.6\\
Fe I&	4994.129	&0.915&	-3.08&	102.0&108.2&	108.9&108.6&	107.2&	109.0&108.5\\
Fe I&	5141.74	&2.42&	-2.23&	90.6	&93.0&	92.9	&93.6&	92.8&	93.3&95.5\\
Fe I&	5198.71	&2.22&	-2.14&	99.2&103.0&	103.8&104.5&	103.5&	104.3&104.7\\
Fe I&	5225.525	&0.11&	-4.789&	74.9&71.1&	71.9&73.7&	71.7	&74.7&76.3\\
Fe I&	5242.49	&3.63&	-0.99&	87.7	&95.9&	96.8&96.9&	95.1&	96.7&96.1\\
Fe I&	5247.05	&0.087&	-4.961&	67.7	&65.7&	65.0&69.8&	69.1&	69.3&70.3\\
Fe I&	5250.208	&0.121&	-4.938&	66.8&62.7&	61.1	&73.0&	66.1	&65.7&73.6\\
Fe I&	5295.31	&4.42&	-1.59&	30.3&37.2&	37.5&36.8&	38.4	&39.3&39.2\\
Fe I&	5322.04	&2.28&	-2.89&	62.3&66.9&	68.2	&66.5&	68.7&	69.4&68.5\\
Fe I&	5373.71	&4.47&	-0.74&	63.6&73.2&	73.5	&72.0&	71.7&	73.9&72.5\\
Fe I&	5379.57	&3.69&	-1.51&	62.4&69.5&	69.5&69.9&	70.9&	71.2&70.2\\
Fe I&	5386.33	&4.15&	-1.67&	33.0&37.0&	38.1&37.5&	39.8	&40.6&39.3\\
Fe I&	5441.34	&4.31&	-1.63&	32.3	&37.9&	38.8&37.9&	39.9	&40.5&39.3\\
Fe I&	5466.396	&4.371&	-0.565&	78.8	&86.4&	89.9&87.0&	87.6&	88.7&86.5\\
\enddata
\tablecomments{This table is available in its entirety in a machine-readable form online. A portion is shown here for guidance regarding its form and content.}
\end{deluxetable}

\begin{deluxetable}{lcccc}
\tabletypesize{\scriptsize}
\tablecolumns{5}
\tablewidth{0pc}
\tablecaption{Stellar Parameters \label{params}}
\tablehead{ 
\colhead{Star} & \colhead{ T$_{\rm{eff}}$ } & \colhead{ log $g$}  &
\colhead{$\rm{[Fe/H]}$} & \colhead{$\xi$ }  \\
\colhead{} & \colhead{(K)} & \colhead{ [cgs] }  &\colhead{(dex)} &\colhead{(km s$^{-1}$)}}
\startdata
NV14 & & & &  \\
\hline
WASP-94A & 6170$\pm$80&4.27$\pm$0.07 &0.26$\pm$0.15 & \nodata \\
WASP-94B  & 6040$\pm$90& 4.26$\pm$0.06 &0.23$\pm$0.14 & \nodata \\
$\Delta$(A-B)  &$+$130$\pm$120 &$+$0.01$\pm$0.09 &$+$0.03$\pm$0.21& \nodata \\
\hline
\hline
This work, Solar Reference & & & &  \\
\hline
WASP-94A & 6198$\pm$8& 4.30$\pm$0.021 &0.318$\pm$0.006 & 1.44$\pm$0.016 \\
WASP-94B  & 6112$\pm$6& 4.38$\pm$0.015& 0.305 $\pm$0.005&1.32$\pm$0.012  \\
$\Delta$(A-B)  &$+$86$\pm$10 & $-$0.08$\pm$0.029& $+$0.013$\pm$0.008&$+$0.012$\pm$0.020  \\
\hline
\hline
This Work, WASP-94B Reference & & & &  \\
\hline
WASP-94A                                & 6198$\pm$4& 4.30$\pm$0.011& 0.319$\pm$0.003& 1.44$\pm$0.008 \\
WASP-94B (same as solar ref.)  & 6112$\pm$6& 4.38$\pm$0.015& 0.305 $\pm$0.005&1.32$\pm$0.012  \\
$\Delta$(A-B)                           & $+$86   $\pm$7       &
$-$0.08$\pm$0.019& $+$0.014$\pm$0.006& $+$0.12$\pm$0.014\\
\hline
\hline
This work, WASP-94B Reference, Isochrone Log $g$ & & & &  \\
\hline
WASP-94A                                & 6194$\pm$5& 4.21$\pm$0.011& 0.320$\pm$0.004& 1.43$\pm$0.008 \\
WASP-94B (log $g$=4.3)  & 6112$\pm$6& 4.30$\pm$0.015& 0.305 $\pm$0.005&1.32$\pm$0.012  \\
$\Delta$(A-B)                           & $+$82   $\pm$7       & $-$0.09$\pm$0.019& $+$0.015$\pm$0.006& $+$0.11$\pm$0.014\\
\enddata
\end{deluxetable}

\begin{deluxetable}{lccccc}
\tabletypesize{\scriptsize}
\tablecolumns{6}
\tablewidth{0pc}
\tablecaption{Derived $\Delta$(A-B) Abundances \label{abuns} }
\tablehead{ 
\colhead{Species} & \colhead{T$_c$} & \multicolumn{2}{c}{\underline{A-B Params,
  log $g_{\rm{B}}=4.3$}} & \multicolumn{2}{c}{\underline{A-B Params, log
  $g_{\rm{B}}=4.3$, no strong lines}} \\
\colhead{} & \colhead{} & \colhead{$\Delta$[X/H]} & \colhead{error} & \colhead{$\Delta$[X/H]} & \colhead{error}\\
\colhead{} & \colhead{(K)} & \colhead{(dex)} & \colhead{(dex)} & \colhead{(dex)} & \colhead{(dex)}}
\startdata
C I& 40     & -0.032  &   0.006 &  -0.032  &   0.006   \\
CH & 40    &  -0.052&  0.015  &    -0.052&  0.015   \\
O I &180    & -0.002 &  0.007   & -0.002$^{1}$ &  0.007 \\
Na I &958   & 0.015 & 0.007   & 0.000  & 0.006 \\
Mg I & 1336&  -0.003&   0.005&  -0.004 & 0.007  \\
Al I &1653  & 0.022 &  0.003  &  0.022 & 0.003 \\
Si I & 1310&  0.020&  0.002  &  0.018 & 0.003 \\
S I &664&-0.021  & 0.007  &  -0.021 &  0.007\\
K I &1006& 0.044 &   0.006 & 0.044$^{1}$   &  0.006\\
Ca I &1517& 0.024 &  0.005  &   0.023  & 0.007  \\
Sc I &1659&  0.007& 0.008   & 0.007   & 0.008 \\
Sc II &1659& 0.012 & 0.006  & 0.008  &  0.006 \\
Ti I &1582&0.004  & 0.005   &  0.003 & 0.005 \\
Ti II &1582&  0.010&   0.006 &   0.011& 0.006 \\
V I &1429& 0.002 & 0.008   &  0.002   & 0.008 \\
Cr I &1296&  -0.011&   0.005 & -0.011  &0.005  \\
Cr II &1296&  0.007&  0.006  &  0.007 &  0.006\\
Mn I &1158& -0.032 &  0.008  &  -0.037 &  0.008\\
Fe I &1334&0.015  & 0.004    &  0.014 &  0.004\\
Fe II &1334& 0.014 &   0.006  & 0.012  &0.006  \\
Co I &1352&  -0.006& 0.007  &  -0.006 & 0.007  \\
Ni I &1353&0.007  &  0.004 &  0.006 & 0.003   \\
Cu I &1037&  -0.031& 0.006  &  -0.031 &  0.005 \\
Zn I &726&  -0.002&    0.007 & -0.002  & 0.007\\
Rb I &800&  0.001&    0.015 &  0.001 & 0.015\\
Y II &1659&  -0.003& 0.008    &  -0.003 &0.009 \\
Zr II &1741& 0.019  & 0.008   &  0.019 &  0.008\\
Ba II &1455& 0.031  &  0.006  & 0.031  & 0.006 \\
\enddata
\tablenotetext{1}{Note: These abundances are based solely on strong
  lines, so strong lines are not culled in these cases.}
\end{deluxetable}

\clearpage


\onecolumn 

\end{document}